\documentclass[aps,llncs,twocolumn]{revtex4-1}
\usepackage{graphicx}
\usepackage{amsmath}
\usepackage{float}

\begin{document}

\title{Spectroscopic signature of two superconducting gaps and their unusual field dependence in RuB$_2$} 

\author{Soumya Datta, Aastha Vasdev, Soumyadip Halder, Jaskaran Singh, Yogesh Singh and Goutam Sheet}

\email{goutam@iisermohali.ac.in}
\affiliation{Department of Physical Sciences, 
Indian Institute of Science Education and Research Mohali, 
 Knowledge City, Sector 81,
Mohali 140306, India}

\begin{abstract}

Recently RuB$_2$ was shown to be a possible two-gap, type-I superconductor. Temperature dependent heat capacity measurements revealed a two-gap superconducting ground state while magnetic field dependent magnetization measurements indicated surprising type-I superconductivity with a very low experimental critical field ($H_c$) $\sim$ 120 Oe. In this paper, we report direct spectroscopic evidence of two superconducting energy gaps in RuB$_2$. We have measured scanning tunnelling spectra exhibiting signature of two gaps on different grains of polycrystalline RuB$_2$, possibly originating from multiple bands. Analysis of the temperature dependent tunnelling spectra revealed that the gaps from different bands evolve differently with temperature before disappearing simultaneously at a single $T_c$. Interestingly, our experiments also reveal that the gaps in quasiparticle density of states survive up to magnetic fields much higher than the bulk $H_c$ and they evolve smoothly with field, unlike what is expected for a type-I superconductor, indicating the existence of a ``mixed state".

\end{abstract}
\maketitle

In 1959, Suhl $et$ $al.$ made an extension to the BCS theory\cite{Bardeen} by considering superconductivity occurring at two different bands crossing the Fermi surface and each band contributing differently to superconductivity, thereby making the superconducting gap ($\Delta$) anisotropic in the momentum space\cite{Suhl}.  Their model also included coupling between the two bands through phonon exchange between quasiparticles belonging to the different bands. They showed that such a superconductor can display two distinct gaps in its quasiparticle excitation spectrum. The amplitude of the two gaps would depend on the strength of the electron-phonon coupling in the two respective bands and on an ``inter-band coupling" term that reflects the possible tunnelling of Cooper pairs between the two bands. Immediately after the extended BCS theory was proposed, an anomaly in the electronic component of the specific heat in V$_3$Si was attributed to multiband nature of superconductivity\cite{Brock}. In 1965, Lung Shen $et$ $al.$ showed from heat capacity measurements that even the simple elemental superconductors like Nb, Ta, V etc., in their purest form, are multiband superconductors\cite{Shen}. However, the research on multiband superconductivity gained momentum in 2001 when two-band superconductivity was discovered\cite{Nagamatshu, Karapetrov, Bouquet, Szabo, Chen, Giubileo} in MgB$_2$ with a remarkably high $T_c$ of 40 K. The two-band nature in case of MgB$_2$ attracted most attention because in this case the signature of the two gaps could be distinctly obtained from a number of experiments including tunnelling spectroscopy where the quasiparticle excitation spectrum directly reflected the two gaps \cite{Giubileo, Schmidt, Iavarone, Gonnelli, Silva-Guillen}. Theoretical models including two BCS gaps with moderate interband scattering \cite{Liu, Bouquet2, Golubov, Choi, Koshelev} nicely described the experimental data on MgB$_2$. This discovery also triggered research on the superconducting properties of a large number of other binary diborides\cite{Vandenberg, Medvedeva}. In this context, OsB$_2$ and RuB$_2$ attracted maximum attention due to their dissimilarities with MgB$_2$ in terms of crystal structure and band structure\cite{Chiodo, Hao, Frotscher}, but seemingly possible multiband superconductivity\cite{SinghY, SinghY2, Bekaert} as in MgB$_2$. OsB$_2$ (Tc $\sim$ 2.1 K) was proposed to be a two-gap superconductor based on heat capacity and penetration depth studies on single crystals\cite{SinghY2}.  In addition, field dependent heat capacity and magnetization measurements suggested that OsB$_2$ is a Type-I superconductor\cite{SinghY, SinghY2}. RuB$_2$ superconducts below 1.5 K \cite{Vandenberg} and the bulk superconducting phase shows surprising ``type I"-like behaviour in terms of its magnetic properties\cite{SinghJ}. The bulk critical field is found to be very low ($\sim$ 120 Oe), the electron-phonon coupling is found to be weak with a coupling constant $\lambda_{ep} \sim$ 0.4 and the temperature dependence of specific heat showed an anomaly that could be fitted well\cite{Yue-Qin, SinghJ} within a two-gap model with the gap values $\Delta_1 \sim$ 0.15 meV and $\Delta_2 \sim$ 0.3 meV. The possibility of multi-gap superconductivity along with type-I character makes RuB$_2$ special. However, in order to obtain a conclusive evidence of multiband superconductivity and to understand the nature of the gaps it is most important to measure the multiple gaps spectroscopically and directly measure the temperature dependence of the gaps.

We have attempted to probe the multiple gaps in RuB$_2$ and their response to temperature and magnetic field by low-temperature scanning tunneling microscopy (STM) and spectroscopy (STS). The polycrystalline samples used for the measurements presented here were from the same batch as that were used for the studies reported in ref. \cite{SinghJ}. The STM and STS experiments were carried out in an ultra-high vacuum (UHV) cryostat working down to 300 mK (Unisoku system with RHK R9 controller). Since STM experiments are extremely sensitive to the surface cleanliness of a material, in order to ensure a pristine surface of RuB$_2$, a few layers of the surface was first removed by mild reverse sputtering in an argon environment $in-situ$ inside an UHV preparation chamber, prior to the STS experiments. All the STM/S experiments were performed with sharp metallic tips of tungsten (W) which were fabricated by electrochemical etching and were cleaned by electron beam bombardment under UHV in the preparation chamber.

\begin{figure}[h!]
	\centering
		\includegraphics[width=.5\textwidth]{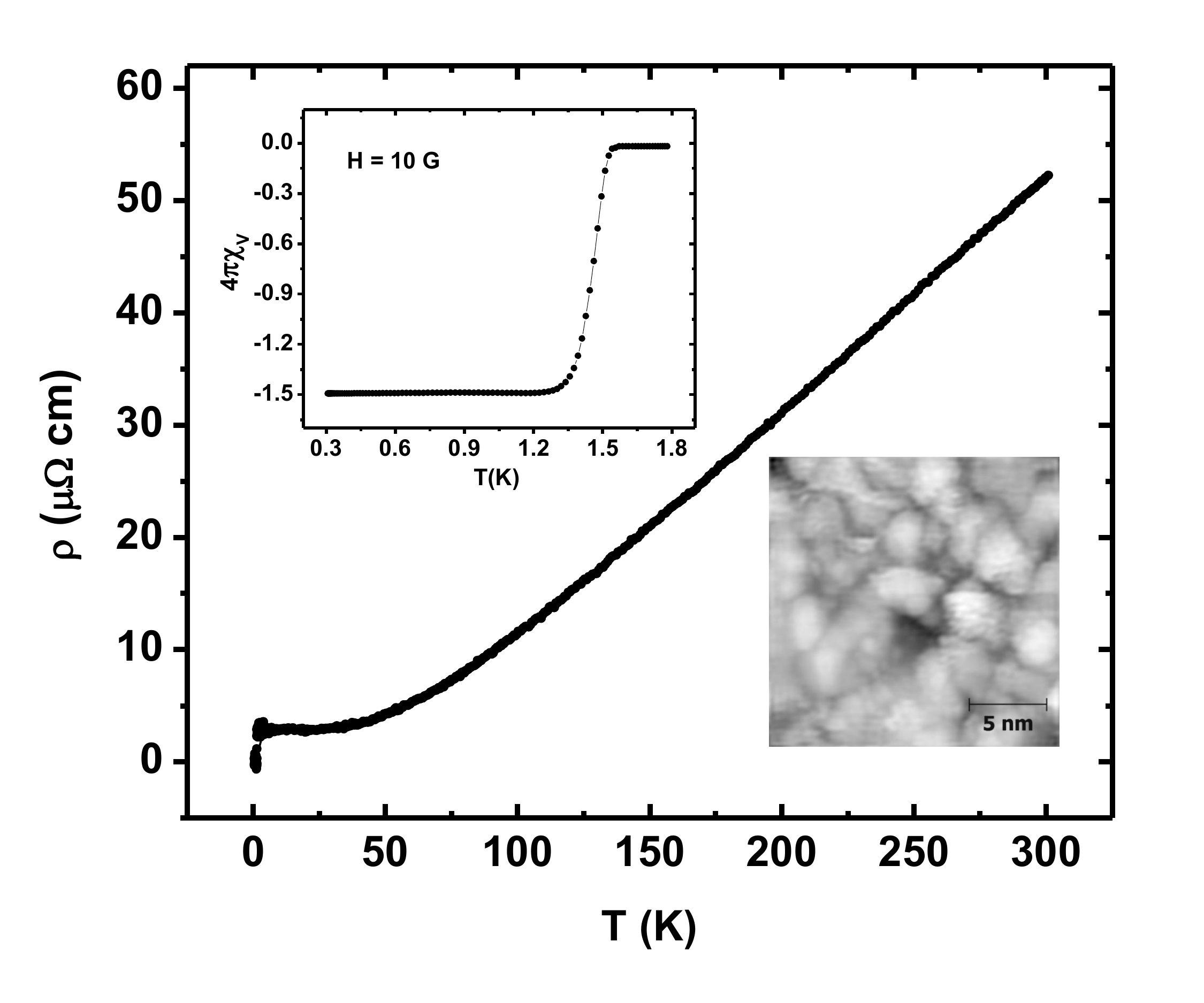}
		\caption{Temperature ($T$) dependence of electrical resistivity ($\rho$) measured in the absence of any magnetic field within the range of $T$ = 0.4 K to 310 K. In the first inset, temperature dependence of dimensionless volume susceptibility $\chi_v$ in terms of the superconducting volume fraction 4$\pi\chi_v$ measured with a magnetic field $H$ = 10 G in zero field cooled (ZFC) condition. Both the data indicate a bulk superconducting transition at $T$ = 1.5 K. In the second inset, STM topographic image of the sample.}

\end{figure}

The superconducting transition in RuB$_2$ was measured by four probe resistivity in a Quantum Design physical properties measurement system (PPMS) and magnetic susceptibility using a He$^3$ insert in a SQUID magnetometer (Cryogenics Limited). Temperature dependence of the electrical resistivity $\rho$ is shown in Fig. 1. Resistivity was measured with an excitation current of 5 mA and at zero applied magnetic field. The temperature dependent volume  magnetic susceptibility $\chi_v$ is also shown in the upper $inset$ of the same Figure. This was measured in the zero-field-cooled (ZFC) state and in presence of a magnetic field of 10 G. Both these measurements confirmed a superconducting transition at 1.5 K. In the lower $inset$ of Fig. 1, we also show an STM topographic image of the RuB$_2$ surface on which reverse sputtering was done for cleaning. The image shows clearly visible grains with $\sim$ 4 nm average grain size. Each grain can be considered as a tiny single crystallite with arbitrary orientation. Hence, by probing different grains it is possible to predominantly probe different crystal facets of RuB$_2$, albeit with no knowledge on the facet being probed.

\begin{figure}[h!]
	\centering
	\includegraphics[width=.5\textwidth]{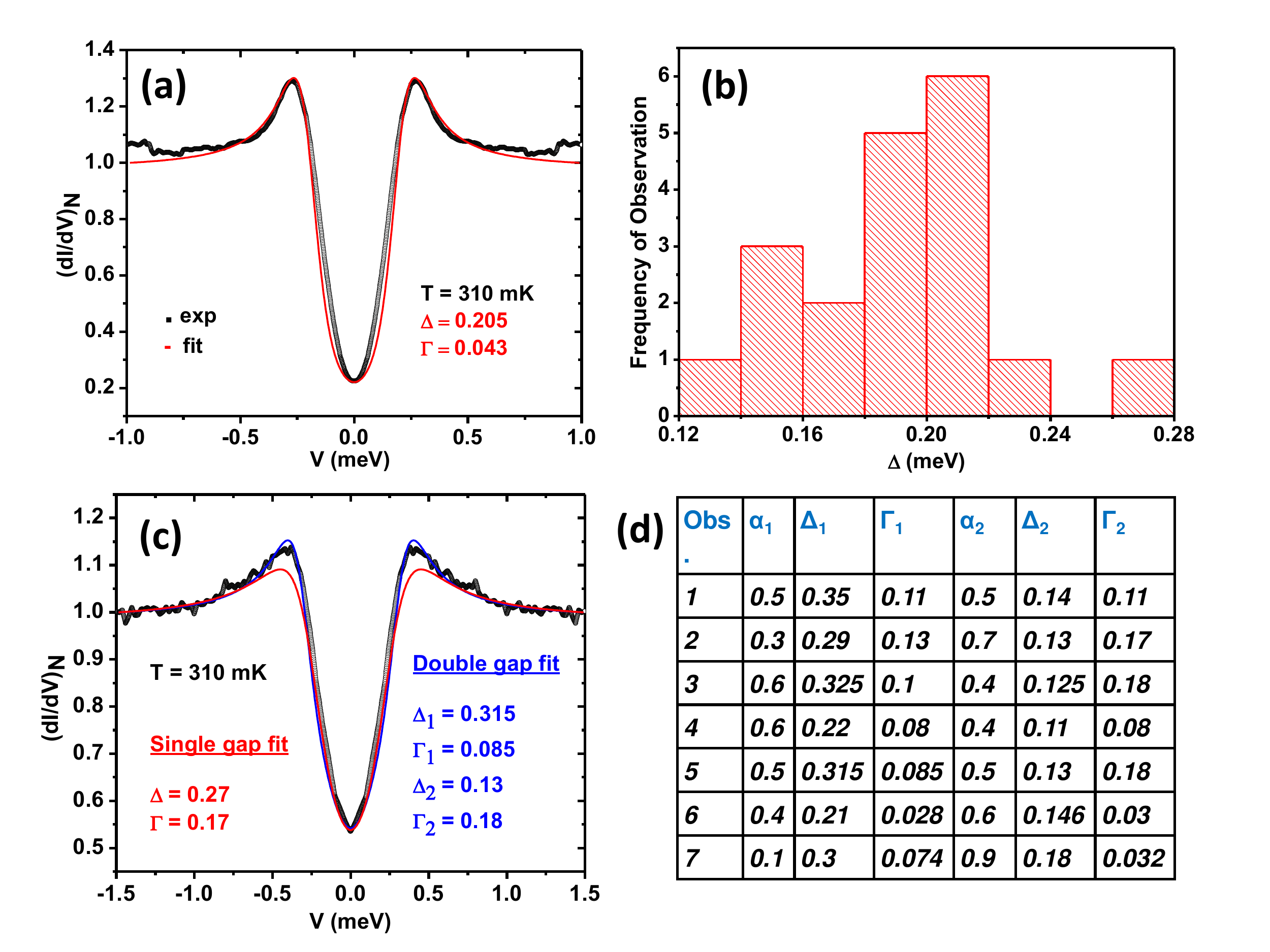}
	\caption{Two representative tunnelling spectra ($dI/dV$ vs $V$ plots) \textbf{(a)} showing existence of a single superconducting energy gap ($\Delta$) and \textbf{(c)} showing double superconducting energy gap. ( $\Delta_1$ and $\Delta_2$ ) with theoretical fits using Dynes equation. For comparison, best possible single gap fit is also shown (red line) in (c). \textbf{(b)} Statistics for all spectra with a single gap Dynes equation fits. \textbf{(d)} Table showing statistics for all spectra with double gaps Dynes equation fits.}
	
\end{figure}

The STS experiments were done by bringing the STM tip on the central parts of different grains and subsequently recording the $dI/dV$ vs $V$ spectra for every such point. In Figs. 2(a) and 2(c) we show two representative tunnelling spectra at lowest temperature. All the spectra are normalised to the conductance at 1.5 mV where they become nearly flat and symmetrised. The spectra were analysed by comparing them with numerically generated spectra using the expression for tunneling current within a single band model given by $I(V) \propto \int_{-\infty}^{+\infty} N_s(E)N_n(E-eV)[f(E)-f(E-eV)]dE$, where $N_s(E)$ and $N_n(E)$ are the normalized DOS of the superconducting sample and the normal metallic tip respectively while $f(E)$ is the Fermi-Dirac distribution function \cite{Bardeen}. Within the single band model, $N_s$ is given by Dyne's formula: $N_s(E) \propto Re\left(\frac{(E-i\Gamma)}{\sqrt{(E-i\Gamma)^2-\Delta^2}}\right)$, where $N_s(E)$ is the density of states at energy $E$, $\Delta$ is the superconducting energy gap and, $\Gamma$ is an effective broadening parameter incorporated to take care of slight broadening of the BCS density of states possibly due to finite quasi-particle life time \cite{Dynes}. The theoretical plot (red curve) is also shown on the experimental spectrum (black line) presented in Figure 2(a). As it is seen, the theoretical curve falls almost completely on the experimental data points, barring very little discrepancy. Here, the $dI/dV$ vs $V$ spectrum shows two clear peaks at $\pm$265 $\mu$V symmetric about $V$ = 0. The position of these coherence peaks, which are signature of the superconducting phase in RuB$_2$, provides a direct measure of the superconducting energy gap ($\Delta$). The extracted value of this gap ($\Delta$) and effective broadening parameter ($\Gamma$) from the theoretical fit are respectively 205 $\mu$eV and 43 $\mu$eV. Similar spectra where single gap fitting was possible were obtained on a number of different grains and they provided a wide range of values for $\Delta$ ranging from 130 $\mu$eV to 280 $\mu$eV, a statistics of which is shown in Fig. 2(b). Despite a large standard deviation of the measured gap values, $\Delta$ follows a central tendency to be around 200 $\mu$eV considering all the spectra that could be fitted using a single gap model.

The spectrum in Fig. 2(c), however, showed deviation from the best possible single band theoretical fit (red curve) prominently around the two peaks. Hence we considered a two gap model \cite{Suhl, Nicol} for this type of spectra. According to the argument of Suhl $et$ $al.$, theoretically, the quasiparticle excitation spectrum for a two band model can be calculated simply by adding the two single gap BCS spectra for the two respective bands (say, band 1 and band 2) \cite{Suhl}. The density of states of the $j$-th band can be written as $N_{s,j} (E) = N_j(E_F) Re\left(\frac {|E-i\Gamma_j|}{\sqrt{(E-i\Gamma_j)^2-\Delta_{0j}^2}}\right)$, where $j$ is the band index, $N_j(E_F)$ is the normal state density of states at the Fermi level corresponding to the $j$th band. $\Delta_{0j}$ is the amplitude of the superconducting energy gap formed in the $j$th band. Now, the tunnelling current will take the form $I(V) \propto \sum_{j = 1,2}\alpha_j\int_{-\infty}^{+\infty} N_{sj}(E)N_n(E-eV)[f(E)-f(E-eV)]dE$, where $\alpha_j$ is the relative contribution of the $j$-th band to the tunneling current. Using this model, the spectra as in Figure 2(c) could be fitted extremely well over the entire energy range (see the blue curve on the black experimental data points). The extracted values of two superconducting gaps are $\Delta_{01}$ = 315 $\mu$eV and $\Delta_{02}$ = 130 $\mu$eV where corresponding two broadening parameters are $\Gamma_1$ = 74 $\mu$eV and $\Gamma_2$ = 180 $\mu$eV. The measured values of the two gaps match well with those obtained from the bulk measurements \cite{SinghJ}. The relative contributions of the two bands to the total tunnelling current are equal, i.e. $\alpha_1$ = $\alpha_2$ = 0.5. The table in Fig. 2(d) describes the relative contributions of two bands ($\alpha_1$ and $\alpha_2$) and values of the corresponding pairs of $\Delta$s and $\Gamma$s in a set of spectra that could be well-fitted by the two band model.

Such variation of the type of spectra where only some of them can be described by a single gap and others not, can be understood if a more realistic two band model\cite{Schopohl, Noat} is considered. In a more realistic two-band model interband quasiparticle scattering terms should also be included in the calculations, particularly for polycrystalline RuB$_2$. If two interband scattering frequencies corresponding to the two bands are $\Gamma_{12}$ and $\Gamma_{21}$ respectively, the gaps in the two respective bands are modified as

 $\Delta_1 (E) = \frac{\Delta_{01} + \Gamma_{12}/\sqrt{(\Delta_2(E))^2-(E-i\Gamma_{21})^2}}{1+\Gamma_{12}/ \sqrt{(\Delta_2(E))^2-(E-i\Gamma_{21})^2}}$,

  $\Delta_2 (E) = \frac{\Delta_{02} + \Gamma_{21}/\sqrt{(\Delta_1(E))^2-(E-i\Gamma_{12})^2}}{1+\Gamma_{21}/ \sqrt{(\Delta_1(E))^2-(E-i\Gamma_{12})^2}}$.

These are two coupled equations reflecting that the gap of one band is expected to be affected by that in the other band as if through a ``proximity effect" in the momentum space. With such modification due to interband scattering, the modified quasiparticle excitation spectrum corresponding to the $j$th band can be written as 

$N_{sj}^{mod}(E) =  N_j(E_F) Re\left(\frac {|E-i\Gamma_j|}{\sqrt{(E-i\Gamma_j)^2-\Delta_{j}^2}}\right)$. 

and the total density of states will be $N_s^{mod} (E)= \sum_{j =1,2}N_{sj}$.

Hence, the shape of the resultant $N_s(E)$ is significantly influenced by not just the gaps in the two bands and the interband tunnelling of Cooper pairs, but also by the interband scattering terms ($\Gamma_{12}$ and $\Gamma_{21}$). These effects may sometimes smear the two distinct peaks in the experimentally measured spectrum depending on which band is predominantly measured\cite{Schopohl}. In complex two band superconductors, where additional effects like $k$-dependence of interband tunnelling can appear, a distribution of spectral shapes is desired\cite{Noat}. When both $\Gamma_{12}$ and $\Gamma_{21}$ are large for a given $k$-direction, the two gaps may seem to merge thereby displaying a single-band like spectrum, like the one shown in Fig. 2(a). This is why most of the times the superconductors, unless in extremely pure form, do not show multi-band nature even if that is present. For such strong interband scattering, the gap may deviate slightly from the BCS-like temperature dependence. For certain $k$-direction, when $\Gamma_{12}$ and $\Gamma_{21}$ are negligibly small, the two gaps may appear distinctly in a single spectrum. On the other hand, for $k$-directions with moderate values of $\Gamma_{12}$ and $\Gamma_{21}$, an overall broadened spectrum can be obtained where single band model will fail though visually the spectrum may look like a single gap BCS one. It seems, the type of spectra represented by the one in Fig. 1(c) was obtained from grains where this condition was satisfied. In this case, the two gaps are also expected to exhibit dramatically different temperature dependence. In order to investigate that, we have carried out detailed temperature dependence of such spectra.

\begin{figure}[h!]
	\centering
	\includegraphics[width=.5\textwidth]{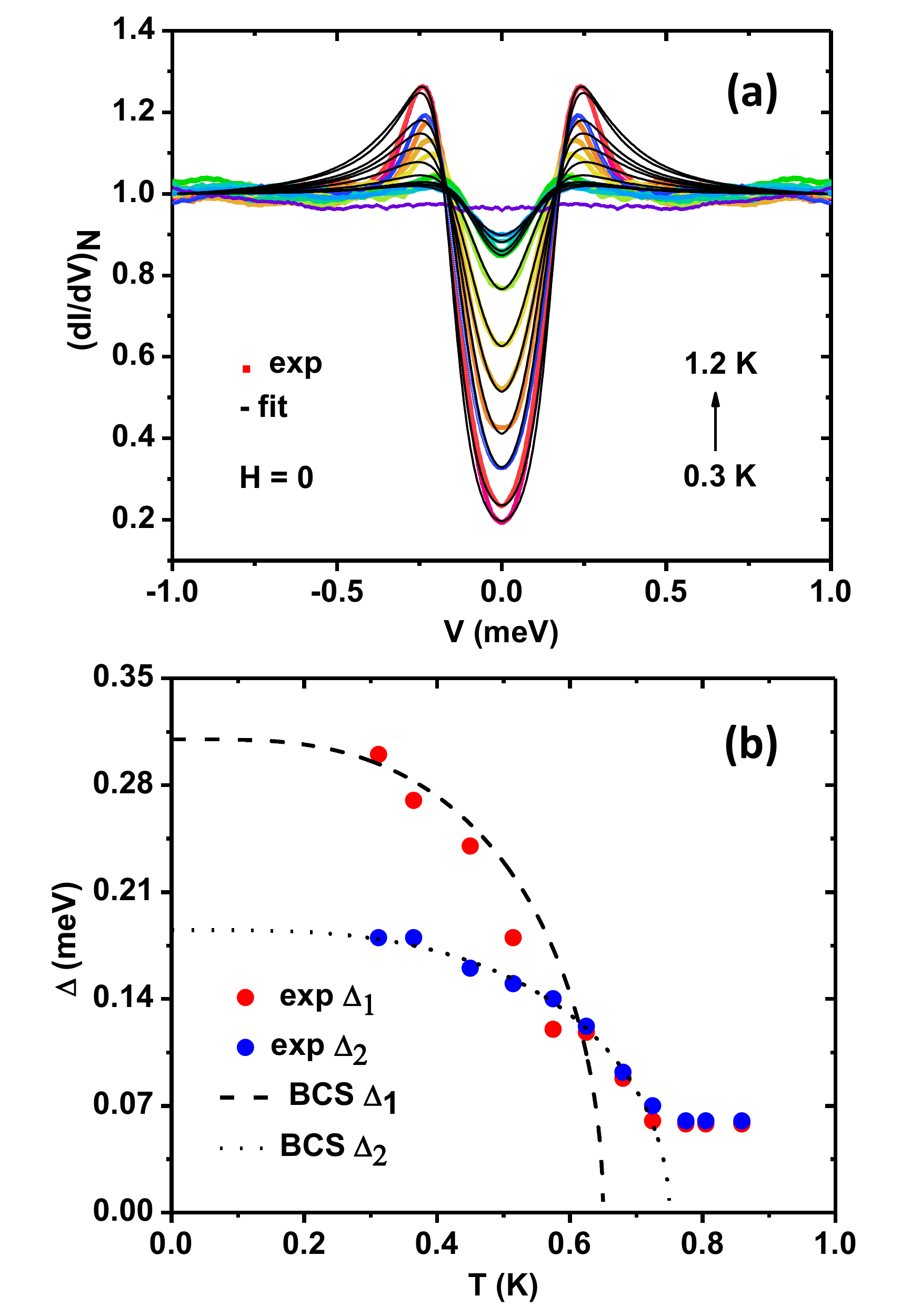}
	\caption{\textbf{(a)} Temperature dependence of tunnelling conductance spectra with theoretical fits considering two gaps in absence of any magnetic field. \textbf{(b)} Evolution of the two gaps ( $\Delta_1$ and $\Delta_2$ ) with temperature, extracted from the plot (a) along with their temperature dependence as per BCS theory.}
	
\end{figure}

In Fig. 3(a) we show the temperature dependence of another such spectrum ( last entry in the table of Fig. 2(d) ) for which two-gap fitting worked. The experimentally obtained spectra are represented by coloured lines where the coherence peaks gradually decrease with increasing temperature and all the features disappear at 1.2 K, close to the bulk $T_c$ of RuB$_2$. The theoretical fits within two-band model are shown on top of the experimental data points as black lines in Fig. 3(a). For the two-band fitting, the values of $\alpha_1$ and $\alpha_2$ were kept unchanged over the entire temperature range. The two gaps extracted from the fits are plotted with temperature in Fig. 3(b) with red ($\Delta_1$) and blue ($\Delta_2$) dots. The two gaps follow different temperature dependence. Around 0.6 K, the two gaps merge with each other, and at that point the larger gap ($\Delta_1$) starts deviating from the BCS line\cite{Bardeen} (the black dashed line) forming a tail that shows a tendency to disappear at a higher temperature. $\Delta_2$ starts deviating from its BCS-like dependence at 0.8 K and beyond this, that also forms a tail. These observations are consistent with the scattering regime\cite{Suhl} where $\Gamma_{12}$ and $\Gamma_{21}$ are moderate and most possibly $\Gamma_{12}$ and $\Gamma_{21}$ also evolve with temperature causing the merging of the gaps at one point.

\begin{figure}[h!]
	\centering
	\includegraphics[width=.5\textwidth]{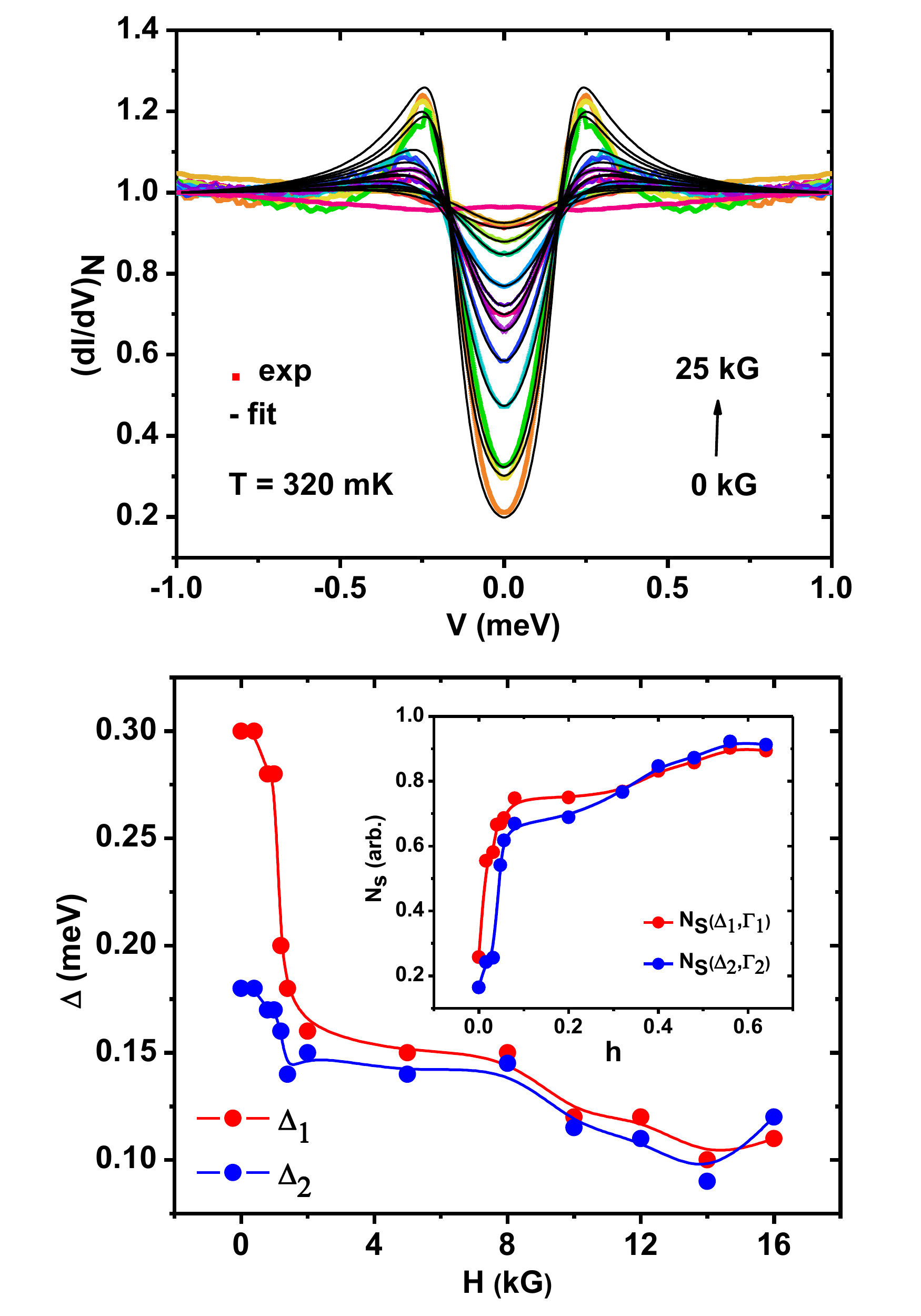}
	\caption{\textbf{(a)} Magnetic field dependence of tunnelling conductance spectra with theoretical fits at $T$ = 0.32 K. \textbf{(b)} Evolution of $\Delta_1$ and $\Delta_2$ with the magnetic field, extracted from the plot (a). $inset$: In arbitrary unit, zero-bias density of states corresponding to two different bands as a function of reduced magnetic field $h$ = $H$/$H_{c(l)}$, where $H_{c(l)}$ = 25 kG is the local critical field}
	
\end{figure}

Now we focus on the magnetic field dependence of the spectra and their two-band analysis. The experimentally obtained field-dependent STS spectra are shown by coloured lines in Fig. 4(a). The corresponding two-band fits are also shown which fall on the experimental data. The evolution of the extracted two gaps with magnetic field are shown in Fig. 4(b). For this set of spectra, the larger gap $\Delta_1\sim$ 300 $\mu$eV. With increasing magnetic field, $\Delta_1$ first falls rapidly and attains 50\% of its zero field value at a magnetic field of 3 kG. In contrast, the smaller gap ($\Delta_2$) shows a small change within this field range showing only 16\% change (from 180 $\mu$eV at H = 0 to 150 $\mu$ev) at H = 3 kG. Beyond this point, the difference between the two gaps become very small and the spectra obtained for magnetic fields larger than 3 kG behaved approximately like single gap spectra. This observation indicates the density of states of the two bands are nearly of similar magnitude at higher magnetic fields when the superconducting order is suppressed. This indicates that the two band superconductivity in this case is not caused by a significant difference in DOS in the two bands. This is not surprising if one looks at the calculated DOS\cite{SinghJ} for the 4$d$ orbital of Ru and the 2$p$ orbital of B, the only two bands that contribute to the Fermi surface, it is seen that there is a very small difference between the DOS near $E_F$ for the two bands. In fact, slightly above $E_F$, the DOS of the two bands become almost identical. Intrinsic defects and electronic disorder may cause a shift in $E_F$ making the DOS in the two bands nearly equal. Hence, the mechanism that might lead to the difference in the two gaps is expected to be a difference in the Eliashberg electron-phonon coupling terms ($\lambda_{11}$ and $\lambda_{22}$) in the respective bands\cite{Liu, Choi2, Nicol}. Qualitatively, a comparison of the Fermi velocities ($v_F$) in the two bands based on the calculation presented in ref. \cite{SinghJ} revealed that one band has smaller $v_F\sim$ 1.1 x 10$^6$ m/s  than the other band ($v_F\sim$ 1.6 x 10$^6$ m/s). The band with smaller $v_F$ is expected to have larger $e-p$ coupling causing the larger gap.

We also note that the local critical field ($H_{c(l)}$) at which the features (like, the gap) associated with superconductivity disappear is more than 1 Tesla which is higher by two orders than the bulk critical field ($H_c\sim$ 100 G) measured earlier\cite{SinghJ}. Furthermore, while the bulk measurements revealed a first order disappearance of superconductivity at $H_c$ as in a type-I superconductor, our measurements indicate that both the superconducting energy gaps survive above the bulk $H_c$ and disappear at a much higher critical field. Intuitively, between the bulk $H_c$ and the local $H_{c(l)}$, the system may be in its mixed state.

Within the theory of a two-band superconductor in the mixed state as developed by Koshelev and Golubov\cite{Koshelev}, a larger vortex core size corresponding to the band forming the smaller gap is the general property of two-band superconductors. This yields two different field scales which means, the band forming the larger gap is expected to attain its normal state density of states at a smaller magnetic field than the field at which the band forming the smaller gap does so. The rapid drop of the larger gap $\Delta_1$ is consistent with this idea. In order to show this effect more clearly, we have calculated the magnetic field dependence of the density of states for the respective bands using the parameters obtained for fitting and by using the formula for $N_{sj}(E)$ described earlier. As it can be seen in the $inset$ of Figure 4, $N_{s1}$ rises up and saturates slightly faster than $N_{s2}$. This further supports the idea that the behaviour of the spectra beyond 3 kG is dictated by a mixed state. To note, all the theoretical work discussed above were done in the context of MgB$_2$. Here, understanding of the origin of the mixed state when the bulk of the system behaves like a type-I superconductor, demands the development of a two-band theory specific to RuB$_2$ in magnetic field.  

To conclude, we have presented detailed STM and STS investigation of two band superconductivity in RuB$_2$. We provided the first direct evidence of two-gap superconductivity in RuB$_2$ confirming previous indications based on indirect bulk measurements. The experiments also indicated the emergence of an unusually large critical field far above the bulk critical field of RuB$_2$ where the bulk superconductivity is destroyed but the local spectral gap continues to remain finite. This observation invites additional RuB$_2$ specific theoretical considerations for better understanding of multiband superconductivity in general.

We thank Shekhar Das for his help with the experiments. We thank Dr. Divya Srivastava for providing us with the information on Fermi velocity of the bands. GS acknowledges financial support from the Department of Science and Technology (DST) through Swarnajayanti fellowship (Grant number: \textbf{DST/SJF/PSA-01/2015-16}).


\begin{thebibliography}{100}
	
	

\bibitem{Bardeen}J. Bardeen, L. N. Cooper, and J. R. Schrieffer,\textit{ Phys. Rev.}\textbf{ 108}, 1175 (1957).


\bibitem{Suhl}H. Suhl, B. T. Matthias, and L. R. Walker,\textit{  Phys. Rev. Lett.}\textbf{ 3}, 552 (1959).


\bibitem{Brock}J. C. F. Brock,\textit{  Solid State Comm.}\textbf{ 7}, 1789 (1969).


\bibitem{Shen}Lawrence Yun Lung Shen, N. M. Senozan, and Norman E. Phillips,\textit{  Phys. Rev. Lett.}\textbf{ 14}, 1025 (1965).


\bibitem{Nagamatshu}Jun Nagamatsu, Norimasa Nakagawa, Takahiro Muranaka, Yuji Zenitani, and Jun Akimitsu,\textit{ Nature}\textbf{ 410}, 63–64 (2001).


\bibitem{Karapetrov}G. Karapetrov, M. Iavarone, W. K. Kwok, G. W. Crabtree, and D. G. Hinks,\textit{ Phys. Rev. Lett.}\textbf{ 86}, 4374 (2001).


\bibitem{Bouquet}F. Bouquet, R. A. Fisher, N. E. Phillips, D. G. Hinks, and J. D. Jorgensen,\textit{ Phys. Rev. Lett.}\textbf{ 87}, 047001 (2001).


\bibitem{Szabo}P. Szabó, P. Samuely, J. Kačmarčík, T. Klein, J. Marcus, D. Fruchart, S. Miraglia, C. Marcenat, and A. G. M. Jansen,\textit{ Phys. Rev. Lett.}\textbf{ 87}, 137005 (2001).


\bibitem{Chen}X. K. Chen, M. J. Konstantinović, J. C. Irwin, D. D. Lawrie, and J. P. Franck,\textit{ Phys. Rev. Lett.}\textbf{ 87}, 157002 (2001).


\bibitem{Giubileo}F. Giubileo, D. Roditchev, W. Sacks, R. Lamy, D. X. Thanh, J. Klein, S. Miraglia, D. Fruchart, J. Marcus, and Ph. Monod,\textit{ Phys. Rev. Lett.}\textbf{ 87}, 177008 (2001).


\bibitem{Schmidt}H. Schmidt, J. F. Zasadzinski, K. E. Gray, and D. G. Hinks,\textit{ Phys. Rev. Lett.}\textbf{ 88},  127002 (2002).


\bibitem{Iavarone}M. Iavarone, G. Karapetrov, A. E. Koshelev, W. K. Kwok, G. W. Crabtree, D. G. Hinks, W. N. Kang, Eun-Mi Choi, Hyun Jung Kim, Hyeong-Jin Kim, and S. I. Lee,\textit{ Phys. Rev. Lett.}\textbf{ 89}, 187002 (2002).


\bibitem{Gonnelli}R. S. Gonnelli, D. Daghero, G. A. Ummarino, V. A. Stepanov, J. Jun, S. M. Kazakov, and J. Karpinski,\textit{ Phys. Rev. Lett.}\textbf{ 89}, 247004 (2002).


\bibitem{Silva-Guillen}J. A. Silva-Guillen, Y. Noat, T. Cren, W. Sacks, E. Canadell, and P. Ordejon,\textit{ Phys. Rev. B}\textbf{ 92}, 064514 (2015).


\bibitem{Liu}Amy Y. Liu, I. I. Mazin, and Jens Kortus,\textit{ Phys. Rev. Lett.}\textbf{ 87}, 087005 (2001).


\bibitem{Bouquet2}F. Bouquet, Y. Wang, R. A. Fisher, D. G. Hinks,
J. D. Jorgensen, A. Junod, and N. E. Phillips,\textit{ Europhys. Lett.}\textbf{ 56}, 6, 856–862 (2001).


\bibitem{Golubov}A A Golubov, J Kortus, O V Dolgov, O Jepsen, Y Kong, O K Andersen, B J Gibson, K Ahn and R K Kremer,\textit{ J. Phys. Condens. Matter }\textbf{ 14}, 1353 (2002).


\bibitem{Choi}Hyoung Joon Choi, David Roundy, Hong Sun, Marvin L. Cohen,
and Steven G. Louie,\textit{ Nature}\textbf{ 418}, 758–760 (2002).


\bibitem{Koshelev}A. E. Koshelev and A. A. Golubov,\textit{ Phys. Rev. Lett.}\textbf{ 90}, 177002 (2003).


\bibitem{Vandenberg}J. M. Vandenberg, B. T. Matthias, E. Corenzwit, and H. Barz,\textit{ Mater. Res. Bull.}\textbf{ 10}, 889 (1975).


\bibitem{Medvedeva}N. I. Medvedeva, A. L. Ivanovskii, J. E. Medvedeva, and A. J. Freeman,\textit{ Phys. Rev. B}\textbf{ 64}, 020502(R) (2001).


\bibitem{Chiodo}S. Chiodo, H.J. Gotsis, N. Russo, E. Sicilia,\textit{ Chem. Phys. Lett.}\textbf{ 425}, 311–314 (2006).


\bibitem{Hao}Xianfeng Hao, Yuanhui Xu, Zhijian Wu, Defeng Zhou, Xiaojuan Liu, and Jian Meng,\textit{ J. Alloys Compounds }\textbf{ 453}, 413–417 (2008).


\bibitem{Frotscher}M. Frotscher, M. Hölzel, and B. Albert,\textit{ Z. anorg. allg. Chem}\textbf{ 636}, 1783-1786 (2010).


\bibitem{SinghY}Yogesh Singh, A. Niazi, M. D. Vannette, R. Prozorov, and D. C. Johnston,\textit{  Phys. Rev. B}\textbf{ 76}, 214510 (2007).


\bibitem{SinghY2}Yogesh Singh, C. Martin, S. L. Bud’ko, A. Ellern, R. Prozorov, and D. C. Johnston,\textit{ Phys. Rev. B}\textbf{ 82}, 144532 (2010).


\bibitem{Bekaert}J. Bekaert, S. Vercauteren, A. Aperis, L. Komendova, R. Prozorov, B. Partoens, and M. V. Milosevi,\textit{ Phys. Rev. B}\textbf{ 94}, 144506 (2016).


\bibitem{Yue-Qin}Wang Yue-Qin, Yuan Lan-Feng, and Yang Jin-Long,\textit{ Chin. Phys. Lett.}\textbf{ 25}, 8, 3036 (2008).


\bibitem{SinghJ}Jaskaran Singh, Anooja Jayaraj, D. Srivastava, S. Gayen, A. Thamizhavel, and Yogesh Singh,\textit{ Phys. Rev. B}\textbf{ 97}, 054506 (2018).


\bibitem{Dynes}R. C. Dynes, V. Narayanamurti, and J. P. Garno,\textit{ Phys. Rev. Lett.}\textbf{ 41}, 1509 (1978).


\bibitem{Nicol}E. J. Nicol and J. P. Carbotte,\textit{ Phys. Rev. B}\textbf{ 71}, 054501 (2005).


\bibitem{Schopohl}N. Schopohl and K. Scharnberg,\textit{ Solid State Comm.}\textbf{ 22}, 371-374 (1977).


\bibitem{Noat}Y. Noat, T. Cren, F. Debontridder, D. Roditchev, W. Sacks, P. Toulemonde, and A. San Miguel,\textit{ Phys. Rev. B}\textbf{ 82}, 014531 (2010).


\bibitem{Choi2}Hyoung Joon Choi, David Roundy, Hong Sun, Marvin L. Cohen, and Steven G. Louie,\textit{ Phys. Rev. B}\textbf{ 66}, 020513(R) (2002).



\end{thebibliography}
\end{document}